%% file: run33_mddm.tex
\documentclass[reprint,aps,superscriptaddress]{revtex4-1}

\usepackage[T1]{fontenc}
\usepackage{amsmath}
\usepackage{amssymb}
\usepackage[tight]{units}
\usepackage{color,graphicx}
\usepackage{soul}

\begin{document}

\input{authors}

\input{body}

\bibliographystyle{h-physrev}
\bibliography{bibliography.bib}

\end{document}

%% file: authors.tex
\newcommand{\mpi}{\affiliation{Max-Planck-Institut f\"ur Physik, D-80805 M\"unchen, Germany}}
\newcommand{\coimbra}{\affiliation{Departamento de Fisica, Universidade de Coimbra, P3004 516 Coimbra, Portugal}}
\newcommand{\vienna}{\affiliation{Institut f\"ur Hochenergiephysik der \"Osterreichischen Akademie der Wissenschaften, A-1050 Wien, Austria \\ and Atominstitut, Vienna University of Technology, A-1020 Wien, Austria}}
\newcommand{\tum}{\affiliation{Physik-Department and Excellence Cluster Universe, Technische Universit\"at M\"unchen, D-85747 Garching, Germany}}
\newcommand{\tuebingen}{\affiliation{Eberhard-Karls-Universit\"at T\"ubingen, D-72076 T\"ubingen, Germany}} 
\newcommand{\oxford}{\affiliation{Department of Physics, University of Oxford, Oxford OX1 3RH, United Kingdom}}
\newcommand{\wmi}{\affiliation{Walther-Mei\ss ner-Institut f\"ur Tieftemperaturforschung, D-85748 Garching, Germany}}
\newcommand{\lngs}{\affiliation{INFN, Laboratori Nazionali del Gran Sasso, I-67010 Assergi, Italy}}

\author{G.~Angloher}
\mpi

\author{A.~Bento}
\coimbra 

\author{C.~Bucci}
\lngs 

\author{L.~Canonica}
\lngs 

\author{X.~Defay}
\tum 

\author{A.~Erb}
\tum
\wmi

\author{F.~v.~Feilitzsch}
\tum 

\author{N.~Ferreiro~Iachellini}
\mpi

\author{P.~Gorla}
\lngs 

\author{A.~G\"utlein}
\vienna

\author{D.~Hauff}
\mpi 

\author{J.~Jochum}
\tuebingen 

\author{M.~Kiefer}
\mpi

\author{H.~Kluck}
\vienna

\author{H.~Kraus}
\oxford

\author{J.-C.~Lanfranchi}
\tum

\author{J.~Loebell}
\tuebingen

\author{A.~M\"unster}
\tum

\author{C.~Pagliarone}
\lngs 

\author{F.~Petricca}
\mpi 

\author{W.~Potzel}
\tum 

\author{F.~Pr\"obst}
\mpi

\author{F.~Reindl}
\mpi

\author{K.~Sch\"affner}
\lngs 

\author{J.~Schieck}
\vienna 

\author{S.~Sch\"onert}
\tum 

\author{W.~Seidel}
\mpi 

\author{L.~Stodolsky}
\mpi 

\author{C.~Strandhagen}
\email{strandhagen@pit.physik.uni-tuebingen.de}
\tuebingen

\author{R.~Strauss}
\mpi 

\author{A.~Tanzke}
\mpi 

\author{H.H.~Trinh~Thi}
\tum 

\author{C.~T\"urko$\breve{\text{g}}$lu}
\vienna

\author{M.~Uffinger}
\tuebingen 

\author{A.~Ulrich}
\tum 

\author{I.~Usherov}
\tuebingen 

\author{S.~Wawoczny}
\tum 

\author{M.~Willers}
\tum

\author{M.~W\"ustrich}
\mpi  

\author{A.~Z\"oller}
\tum

%% file: body.tex

\title{Limits on momentum-dependent asymmetric dark matter with CRESST-II}

\begin{abstract}
The usual assumption in direct dark matter searches is to only consider the spin-dependent or spin-independent scattering of dark matter particles. However, especially in models with light dark matter particles $\mathcal{O}(\mathrm{GeV/c^2})$, operators which carry additional powers of the momentum transfer $q^2$ can become dominant. One such model based on asymmetric dark matter has been invoked to overcome discrepancies in helioseismology and an indication was found for a particle with preferred mass of \unit[3]{$\mathrm{GeV/c^2}$} and cross section of \unit[$10^{-37}$]{$\mathrm{cm^2}$}. Recent data from the CRESST-II experiment, which uses cryogenic detectors based on $\mathrm{CaWO_4}$ to search for nuclear recoils induced by dark matter particles, are used to constrain these momentum-dependent models. The low energy threshold of \unit[307]{eV} for nuclear recoils of the detector used, allows us to rule out the proposed best fit value above.
\end{abstract}

\maketitle

\section{Introduction}
Today, the existence of dark matter in the universe is well established. The nature of this dark matter however remains unresolved. One possible solution would be the existence of new particles beyond the Standard Model. Among the most favored particle candidates are the so-called WIMPs (weakly interacting massive particles). They are created thermally in the early universe and could be detected via elastic scattering off nuclei, causing recoils of a few keV in energy. The WIMP mass range spans from \unit[$\sim 2$]{$\mathrm{GeV/c^2}$} up to \unit[$\sim$]{$\mathrm{TeV/c^2}$}. Asymmetric dark matter (ADM) models offer an alternative mechanism to create the dark matter, connecting the asymmetry observed in the baryonic sector with the dark sector \cite{ReviewAsymmetricDM}. In ADM models the dark matter particles are expected to have a mass of  $\mathcal{O}(\mathrm{GeV/c^2})$.

In both cases the dominant contribution to the interaction between dark matter and normal matter is usually assumed to be either spin-independent (SI) or spin-dependent (SD). There exist however viable models where these channels are suppressed and the dominant interaction becomes dependent on the transferred momentum $q^2$ \cite{MDDM}. Recently it has been shown that momentum-dependent asymmetric dark matter can resolve a disagreement between helioseismological data and solar models resulting in a preferred dark matter mass of \unit[3]{$\mathrm{GeV/c^2}$} and cross section of \unit[$10^{-37}$]{$\mathrm{cm^2}$} \cite{MDDMinSun}.

The CRESST-II experiment, using cryogenic detectors particularly sensitive to light dark matter, is well suited to test these models. The detailed setup of CRESST-II is described in \cite{Angloher2009_run30, Angloher2005_cresstIIproof}. It uses scintillating $\mathrm{CaWO_4}$ crystals as absorber material operated as cryogenic detectors at milli-Kelvin temperatures. Each crystal is equipped with a tungsten transition edge sensor (TES) which measures the phonons induced in a particle interaction. This phonon signal allows a precise reconstruction of the deposited energy independent of the particle type. The scintillation light produced in an interaction is detected in a second absorber made of silicon and sapphire. This so-called light detector is again equipped with a tungsten TES and records the light signal which can be used to discriminate the interacting particle. This is possible since the ratio of light to phonon signal (light yield) depends on the interaction type. Due to their higher ionization density, the light production for nuclear recoils (e.g. dark matter particles or neutrons) is quenched relative to electronic recoils (e.g. gammas or betas).

The most crucial parameter in the search for $\mathcal{O}(\mathrm{GeV/c^2})$ dark matter particles is the energy threshold for nuclear recoils. Here current CRESST detectors feature among the lowest thresholds in the field (as low as \unit[$307 \pm 3$]{eV} \cite{LisePaper}). Another advantage of CRESST detectors for low mass dark matter searches are the relatively light targets oxygen ($\mathrm{A} \approx 16$ ) and calcium ($\mathrm{A} \approx 40$ ) where more energy is transferred in the scattering process compared to heavier targets like germanium or xenon. Together, this leads to the currently best sensitivity to spin-independent dark matter scattering for masses below \unit[$\sim 2$]{$\mathrm{GeV/c^2}$} \cite{LisePaper, TUM40paper}.

\section{Data Set and Analysis} \label{sec:Analysis}
The detector used in this work is described in \cite{LisePaper}. It features the lowest energy threshold (\unit[307]{eV}) of all detectors which makes it best suited for this analysis.

\begin{figure}[tpb]
  \includegraphics[width=\linewidth]{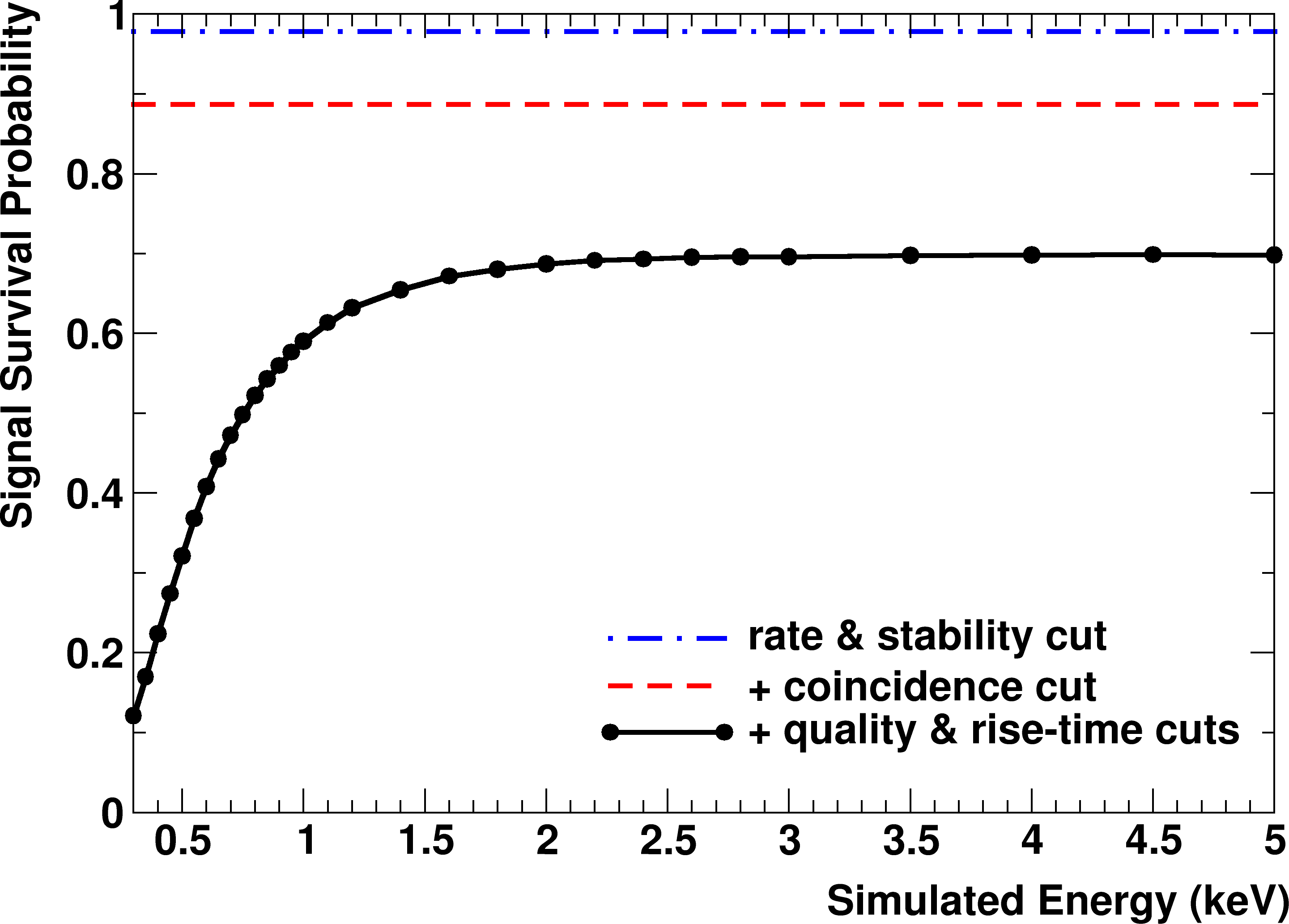}
  \caption
  {Energy dependent signal survival probability after succesive application of the different selection criteria. Figure adapted from \cite{LisePaper}.
}
\label{fig:acceptance}
\end{figure}

The data set for this work contains an exposure of 52.2 kg live days. The data as well as the data reconstruction are the same as in \cite{LisePaper}. The energy reconstruction starts with obtaining the pulse height by fitting a signal template to the recorded pulses. The signal templates are created by averaging a large number of pulses from the \unit[122]{keV} line in a dedicated calibration with a $\mathrm{^{57}Co}$-source. Pulses injected into the TES with a heater are then used to linearize the detector response. Finally, the energy scale is set by the response to $\gamma$-rays of the \unit[122]{keV} peak from the $\mathrm{^{57}Co}$ calibration source.

\begin{figure}[tpb]
  \centering
  \includegraphics[width=0.94\linewidth]{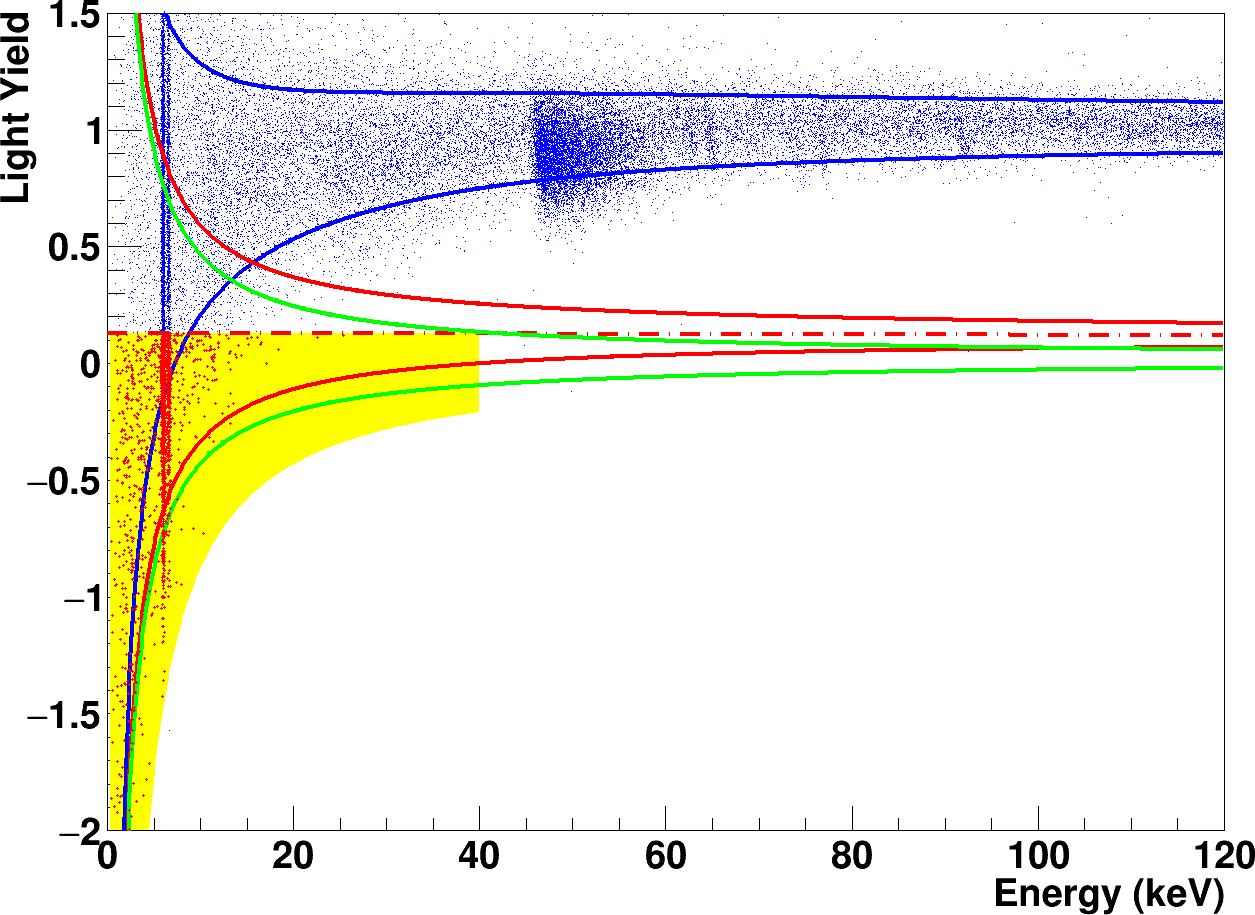}
  \caption
  {Data plotted in light yield-energy plane. The upper and lower 90\% contours of the $\mathrm{e^-/\gamma}$-, oxygen- and tungsten-bands are shown in blue, red and green respectively. The dashed red line indicates the center of the oxygen band which is used as upper boundary for the acceptance region which is shaded in yellow. All events in the acceptance region are highlighted in red. Figure adapted from \cite{LisePaper}.
}
\label{fig:2dplot}
\end{figure}

All data quality cuts are defined on a statistically insignificant training set and blindly applied to the final data set. After the removal of periods where the detector is not in its nominal operating point, events coincident with the muon veto and/or any other detector are removed. Finally, data quality cuts are applied to remove unwanted pulses (e.g. electronic artifacts, pile-up). The signal survival probability of these cuts is determined by applying the cuts on artificial nuclear recoil events which are generated by superimposing the signal template - scaled to the desired energy - and randomly sampled baselines. The resulting energy-dependent signal survival probability (see Fig.~\ref{fig:acceptance}) and the events surviving all cuts (see Fig.~\ref{fig:2dplot}) are identical to the ones in \cite{LisePaper}.

A likelihood fit to this data yields the band containing the $\mathrm{e^-/\gamma}$-events. Two noticeable features stand out in this $\mathrm{e^-/\gamma}$-band: a $\beta$-decay spectrum (starting from \unit[46.5]{keV}) from the intrinsic contamination of the crystal with $\mathrm{^{210}Pb}$ and the double peak around \unit[6]{keV} from an accidental illumination with an $\mathrm{^{55}Fe}$-source deployed to calibrate the light detector of another module. The contamination due to the $\mathrm{^{55}Fe}$-source however has only negligible impact for masses \unit[$\lesssim~5$]{$\mathrm{GeV/c^2}$} since the energy is far above the expected recoil energies.  

The precise amount of reduction of the light output for nuclear recoils off the different target nuclei with respect to the light produced for $\mathrm{e^-}$ or $\gamma$ events is quantified by the so-called quenching factors. They have been precisely measured in an external setup \cite{Strauss2014} and are used to define bands in the energy-light yield plane where we expect nuclear recoils to show up. The validity of this method has been checked by irradiating the detectors with neutrons from an Am-Be calibration source.

The acceptance region for dark matter candidates is defined analogously to \cite{LisePaper} to contain the lower half of the oxygen band bounded below by the lower 99.5\% contour of the tungsten band. The energy range starts at the trigger threshold of \unit[307]{eV}, which is measured with pulses injected into the TES heater (see Fig.~1 of Ref.~\cite{LisePaper}), and extends up to \unit[40]{keV} (yellow shaded region in Fig.~\ref{fig:2dplot}). All events in this region (highlighted in red in Fig.~\ref{fig:2dplot}) are considered as dark matter candidates, no background was subtracted.

The ability to separate $\mathrm{e^-/\gamma}$-events from nuclear recoils depends mainly on the resolution of the light detector which dominates the width of the bands. The modest performance of the light detector used in this module leads to leakage of $\mathrm{e^-/\gamma}$-events into the acceptance region also at higher energies.

\section{Results, Discussion and Outlook}

\begin{figure}[t]
  \centering
  \includegraphics[width=0.85\linewidth]{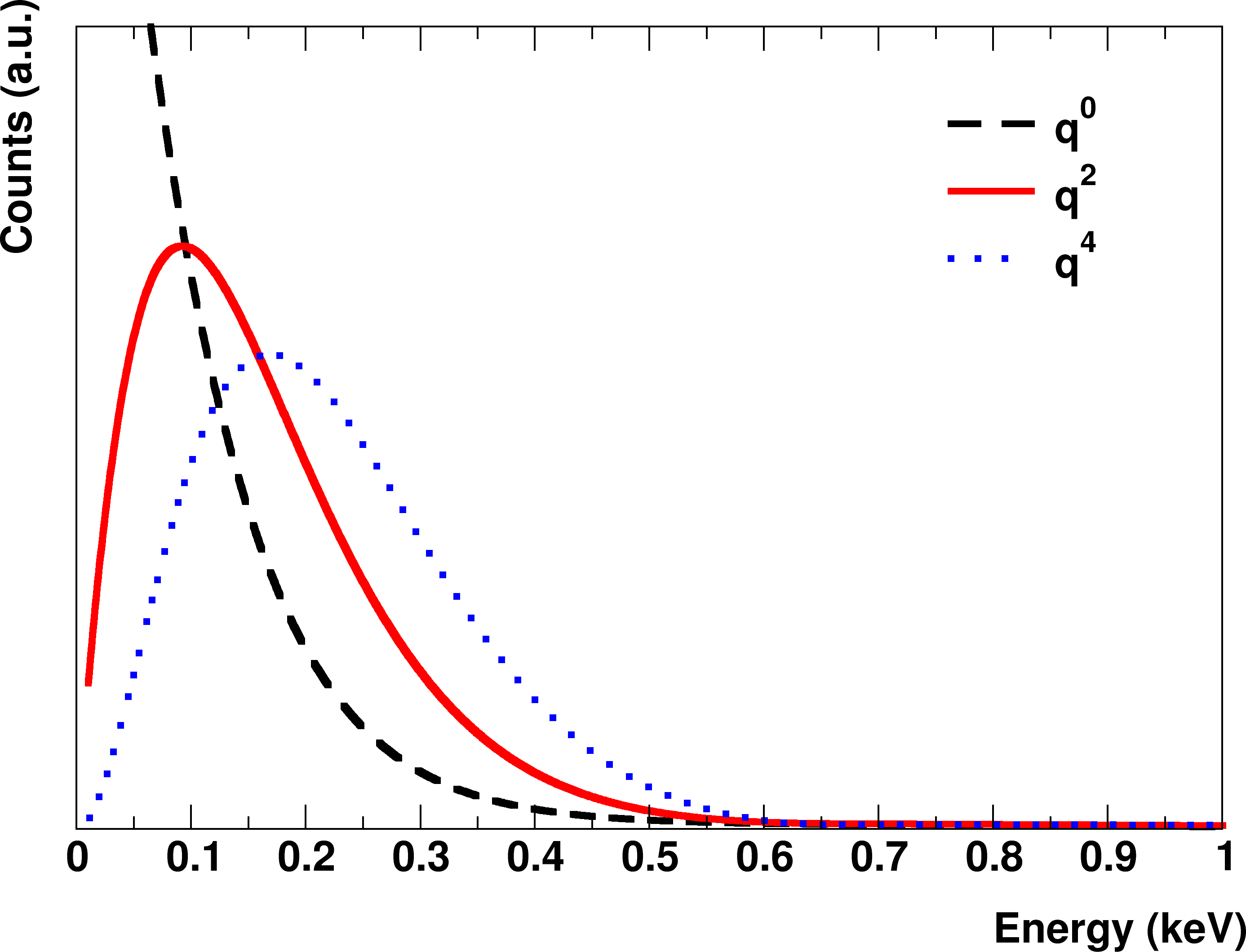}
  \caption
  {Recoil spectra for \unit[3]{$\mathrm{GeV/c^2}$} dark matter particles scattering off $\mathrm{CaWO_4}$. In dashed black the default spin-independent spectrum is shown, in solid red and dotted blue the effect of the different powers of $q$ on the shape of the recoil spectrum can be seen. All spectra are normalized to one, note however that especially for small masses the absolute count rate of the modified spectra may be suppressed by several orders of magnitude.
}
\label{fig:spectra}
\end{figure}

Using the optimum interval method \cite{Yellin2002_Limit,optimum_I}, an upper limit with \unit[90]{\%} confidence level is set on the elastic spin-independent interaction cross section of dark matter particles with nucleons. The dark matter-nucleon cross section is modified \cite{MDDM} by a factor of 

\begin{equation}
\sigma_{\chi-n} = \sigma_0 \left(\frac{q^2}{q_0^2}\right)
\end{equation}

where $q^2$ is the transferred momentum and $q_0$ is a normalization factor chosen to be \unit[40]{MeV} to be consistent with \cite{MDDMinSun}.

This modification of the default spin-independent recoil spectrum causes a suppression of the event rate at lowest energies and leads to a peaked energy spectrum (see Fig.~\ref{fig:spectra}).

To be able to directly compare the results, the halo parameters from \cite{MDDMinSun} are adopted: Maxwell-Boltzmann halo with velocity dispersion of \unit[270]{$\mathrm{km~s^{-1}}$}, velocity of the sun of \unit[220]{$\mathrm{km~s^{-1}}$} and local dark matter density of \unit[0.38]{$\mathrm{GeV~cm^{-3}}$}. The galactic escape velocity is taken as \unit[544]{$\mathrm{km~s^{-1}}$}. These differ slightly from the ones used in \cite{LisePaper}, which however has no significant impact on the results. We use the Helm parametrization of the nuclear form factors to account for deviations from the $\mathrm{A^2}$-dependence of the scattering cross section due to loss of coherence. This approach is valid, since the nuclear physics involved in the $q^2$-dependent scattering is the same as in the standard spin-independent scattering \cite{PhysRevC.89.065501}.

\begin{figure}[t]
  \centering
  \includegraphics[width=\linewidth]{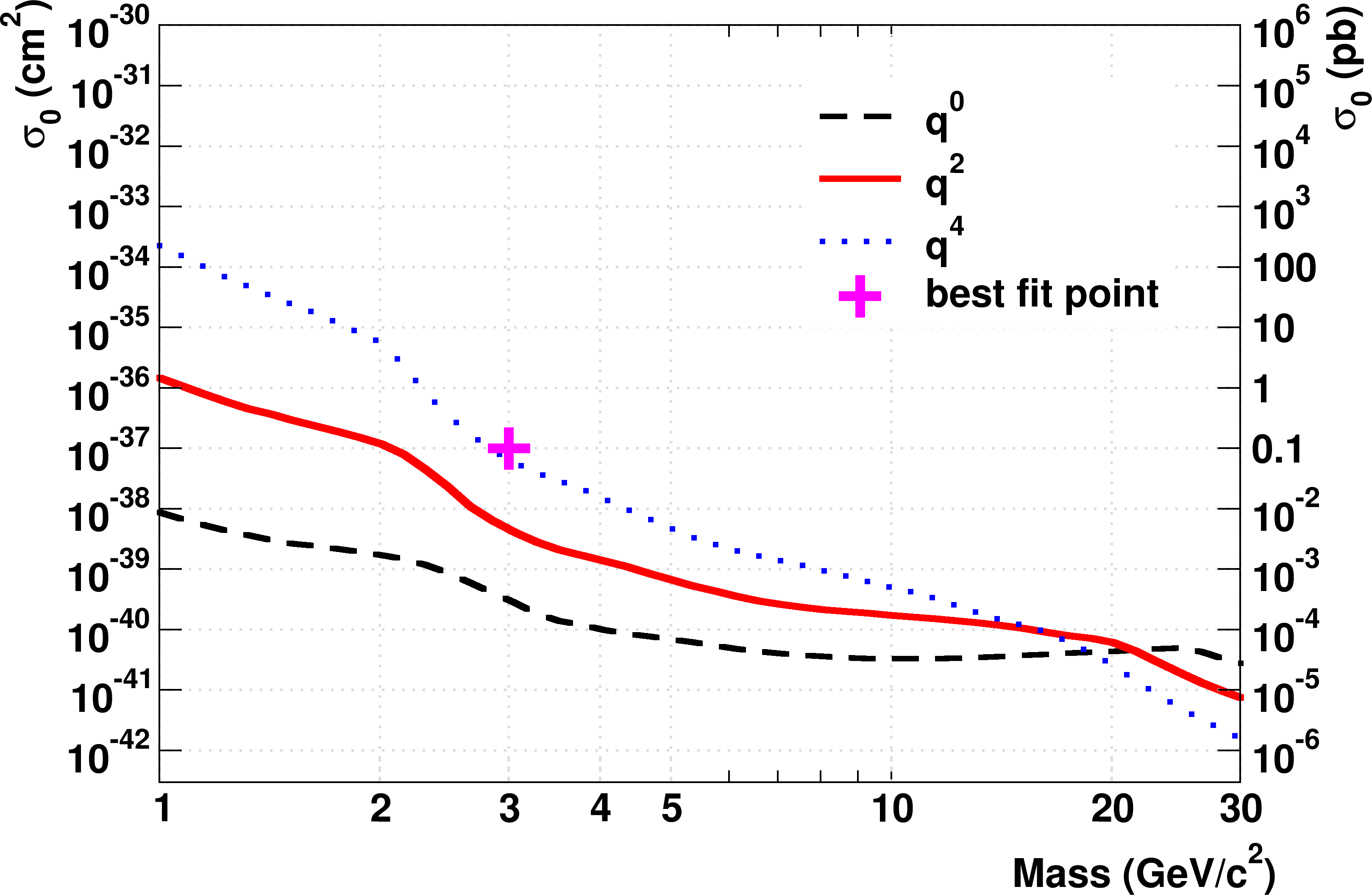}
  \caption
  {\unit[90]{\%} C.L. upper limits on $\sigma_0$ for different for different powers of $q$. The limit for $q^2$-dependent scattering is drawn in solid red ruling out the best fit point from \cite{MDDMinSun} (magenta cross). For comparison also the limit for $q^4$-dependent scattering (dotted blue) and scalar interaction (dashed black) are shown.}
\label{fig:exclusion}
\end{figure}

The resulting exclusion limit for $q^2$-dependent dark matter is drawn in solid red in Fig.~\ref{fig:exclusion}. For comparison also the default SI exclusion (dashed black) and the limit for $q^4$-dependent dark matter (dotted blue) are shown. Our result for $q^2$-dependent scattering excludes the best fit point from \cite{MDDMinSun} for $q^2$-dependent dark matter by an order of magnitude, ruling out this particular model.

The kinks around \unit[2-3]{$\mathrm{GeV/c^2}$} in the exclusion curves are caused by the presence of the different target nuclei in the detector. Above these kinks the scattering is dominated by tungsten due to the expected $\mathrm{A^2}$-dependence of the scattering cross section. Below, the kinetic energy of the dark matter particles is insufficient to cause tungsten recoils above the energy threshold and only oxygen recoils can still be observed at these low masses.

The rather large number of $\mathrm{e^-/\gamma}$-events leaking into the acceptance region limits the sensitivity that can be reached with this detector. For the future it is planned to reduce the dimensions of both the absorber crystal and the light detector. This should lead to an even lower energy threshold (\unit[$\lesssim 100$]{eV}) and improve the discrimination of signal and background events due to enhanced sensitivity in the light channel. In addition, absorber crystals with significantly lower intrinsic background will have an immediate impact on the sensitivity of the detectors \cite{strauss_betagamma_2015}.  

In summary, we have shown that current CRESST detectors are a valuable tool for constraining also more general dark matter models. The low energy thresholds which can be achieved with these detectors provide an unique opportunity to search for light dark matter particles.

\section*{Acknowledgements}

This work was supported by funds of the German Federal Ministry of Science and Education (BMBF),
the Munich Cluster of Excellence (Origin and Structure of the
Universe), the Maier-Leibnitz-La\-bo\-ra\-to\-ri\-um (Garching), the Science
and Technology Facilities Council (STFC) UK, as well as the Helmholtz Alliance for Astroparticle Physics. We gratefully acknowledge the work of Michael Stanger from the crystal laboratory of the TU Munich. We are grateful to LNGS for their generous support of CRESST, in particular to Marco Guetti for his constant assistance.